# The effect of magnetic field and disorders on the electronic transport in graphene nanoribbons

S. Bala Kumar[1*], M. B. A. Jalil[1], S. G. Tan[2], and Gengchiau Liang[1*]

[1] Department of Electrical and Computer Engineering, National University of Singapore, Singapore, 117576,
[2] Data Storage Institute, (A*STAR) Agency for Science, Technology and Research, DSI Building, 5 Engineering Drive 1, Singapore 117608

[*] Corresponding author. E-mail: brajahari@gmail.com and elelg@nus.edu.sg

## **Abstract**

We developed a unified mesoscopic transport model for graphene nanoribbons, which combines the non-equilibrium Green's function (NEGF) formalism with the real-space π-orbital model. Based on this model, we probe the spatial distributions of electrons under a magnetic field, in order to obtain insights into the various signature Hall effects in disordered armchair graphene nanoribbons (AGNR). In the presence of a uniform perpendicular magnetic field ($B^{\perp}$-field), a perfect AGNR shows three distinct spatial current profiles at equilibrium, depending on its width. Under non-equilibrium conditions (i.e. in the presence of an applied bias), the net electron flow is restricted to the edges and occurs in opposite directions depending on whether the Fermi level lies within the valence or conduction band. For electrons at energy level below the conduction window, the $B^{\perp}$-field gives rise to local electron flux circulation, although the global flux is zero. Our study also reveals the suppression of electron backscattering as a result of the edge transport which is induced by the $B^{\perp}$-field. This phenomenon can potentially mitigate the undesired effects of disorders, such as the bulk and edge vacancies, on the transport properties of AGNR. Lastly, we show that the effect of $B^{\perp}$-field on electronic transport is less significant in the multimode compared to the single mode electron transport.

**Introduction**

Graphene, a two-dimensional graphite sheet, has been studied extensively [1-3] due to its unique electronic properties, such as extremely high carrier mobility [4], long spin relaxation length[5], the fractional quantum hall effect [6], Andreev reflection [7] and Klein tunneling [8]. However, graphene is a zero band gap semiconductor where the valence and conduction bands touch at the Brillouin zone corners [1-3]. Due to the zero band gap, it is difficult to implement graphene-based electronic device applications. One of the promising methods to increase the band gap of graphene is by constricting its dimensions, i.e. creating one dimensional structure such as graphene nanoribbons (GNRs) [9-11]. The band gap of a GNR is determined by the quantum confinement along the transverse direction, edge effects and edge patterns. Thus, the energy band gap of GNRs can be varied by modulating the width and the edge profile of GNRs. The two typical types of GNRs, classified based on the edge profiles are: 1) zig-zag GNR (ZGNR), and 2) armchair GNR (AGNR) [9-11] .

Furthermore, several theoretical studies have shown that the electronic and transport properties of the AGNR can be modified by the applying an external magnetic field [12-16]. At moderate magnetic field, the lowest sub-band $|n|=1$ shifts towards E=0, while the other sub-bands shift away from E=0. The modification of the band structures due to magnetic field also gives rise to the Hall Effect [17]. As the magnetic field increases, the cyclotron frequency and the effective mass increases, thus flattening the dispersion profile and forming a Landau level (LL). The forward and backward transport states shift to the opposing edges, decreasing the spatial overlap of these states. When the magnetic field is sufficiently high, such that the electron's cyclotron radius is smaller than the GNR width, LLs composing of dispersionless and parabolic energy bands are formed. The formation of LLs gives rise to the quantum Hall effect (QHE), [18-21] whereby edge states are formed and the electrons are transported only along the edges of the structure. This eliminates the spatial overlap of the forward and backward transport states, resulting in a complete suppression of the backscattering due to disorders in the structure.[17]

Although the electronic and transport properties in GNR under magnetic field have been extensively studied,[12-16,18-21] there are not many studies done on the local features of electron transport in GNR. The previous work done by L. P. Zabro et. al.[22] studied the spatial distribution of nonequilibrium current and the effect of impurities in ZGNRs, while K. Wakabayashi *et. al.*[15] showed the spatial distribution of the equilibrium current in GNRs under magnetic field. However - 1) the local features of the nonequilibrium electron transport in GNRs under magnetic field; and 2) the effect of magnetic field on the backscattering caused by disorders e.g. vacancies – are yet to be well-understood.

Therefore, in this paper, we carry out a detailed study on the effect of magnetic field and backscattering caused by the disorders, on the spatial distribution of the local electron transport in AGNR. Firstly, we study the spatial profile of the equilibrium current in AGNR structure under an external uniform perpendicular magnetic field ($B^{\perp}$-field). We find that, depending on the width of the AGNR, there are three distinct spatial distributions of current under the $B^{\perp}$-field. These results are consistent with the results obtained previously.[15] We further study the spatial distribution of the nonequilibrium current under $B^{\perp}$-field (i.e. in the presence of a finite bias), and compare the current profile for different electron energies: 1) at the Fermi level in conduction band, 2) within the conduction band but below the conduction window, and 3) at the Fermi level in the valence band. We find that, while the net current flows in only one of the edges, there are circulating currents along both the edges due to the Hall effects. Next, we investigate the effect of atomic disorders and backscattering on the transport properties of the AGNR. The effect of different types of disorders on the conductance and spatial current profile are analyzed. Finally, we also investigate the effect of $B^\perp$-field on the electron transport in the disordered structures.

**Methodology.**

The nearest neighbor π-orbital tight binding model [23,24] was employed to investigate the electronic structure of an AGNR under a uniform perpendicular magnetic field ($B^\perp$-field). The Hamiltonian of a GNR is given by

$$H = \sum_{mn} \left( \langle t | m \rangle \langle n | \right) \tag{1}$$

where the term in the brackets in non-zero only for the nearest neighbor atoms and is zero for i=j. When a perpendicular magnetic field ($B^\perp$-field) is applied, the magnetic flux passing through each hexagonal ring of the honeycomb carbon structure, in the unit of flux quantum, $\phi_0$=h/q, is given by $\phi = SB_z / \phi_0$, where S is the area of the hexagon. The flux $\phi$ is used to characterize the strength of the $B^\perp$-field. The magnetic field of $\vec{B} = (0,0,B_z)$ induces a vector potential of $\vec{A} = (-B_z y, 0, 0)$ which satisfies $\nabla \times \vec{A} = \vec{B}$. Under an applied $B^\perp$-field, following

Peierls phase approximation [25], the hopping energy, $t_{n,m}$ acquires a phase, i.e. $t_{n,m}(\phi) = t_{n,m}(0)\exp\left(i\frac{q}{h}\int_{l_m}^{l_n} A(\phi)d\vec{l}\right)$, where $l_{n(m)}$ is the coordinate of atom $n$ ($m$), and $t_{n,m}(0) \approx 3eV$ is the nearest neighbor hopping parameter [26] energy under zero $B^{\perp}$-field.

The electron transport behaviors of the AGNR are studied using the non-equilibrium Green's function (NEGF) formalism [27]. Based on the NEGF formalism, at T=0K, the linear response conductance, g(E, $\phi$) across the GNR can be approximated as

$$g(E_F,\phi) = g_0 Tr\left[\Gamma_S(E_F,\phi)G^r(E_F,\phi)\Gamma_D(E_F,\phi)G^r(E_F,\phi)^+\right] \quad (2)$$

where $G^r(E,\phi) = [EI - H(\phi) - \Sigma_S(E,\phi) - \Sigma_D(E,\phi)]^{-1}$ is the retarded Green's function of the GNR channel, $\Gamma_{S(D)}(E,\phi) = i[\Sigma_{S(D)}(E,\phi) - \Sigma^+_{S(D)}(E,\phi)]$ denotes the coupling between the source (drain) contacts to the GNR, $\Sigma_{S(D)}(E,\phi)$ is the self energy of the left (right) contacts, and $g_0 = e^2/\hbar$ is the quantum conductance. For a given energy E, the electron flux (e-flux) across two atoms n and m is

$$j_{nm}(E) = j_0[t_{mn}G^<_{nm}(E) - t_{nm}G^<_{mn}(E)], \quad (3)$$

where $G^<_{nm}(E)$ is the lesser Greens function [28,29] and $j_0=g_0/e$. The average longitudinal e-flux across dimer N of the GNR and at energy E is represented by $j_x(N,E)$. Note that the e-flux is proportional to current, and direction of e-flux is opposite to the direction of current, since the charge of electron is negative.

# Results and Discussion.

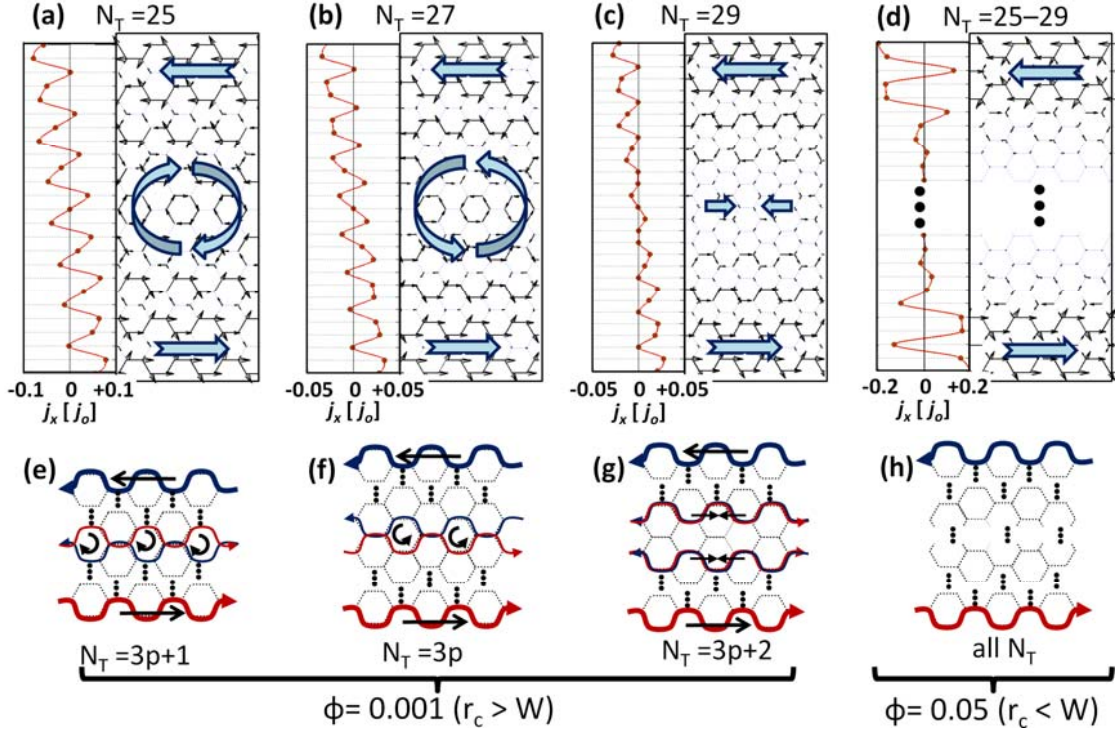

Figure 1. Spatial variation of equilibrium e-flux in AGNRs under $B^\perp$-field. (a), (b), and (c) show the structure of AGNRs with total number of dimers, $N_T$=25, 27, and 29 respectively. The bond e-flux in between two atom, $j_{mn}(E)$ is represented by the arrows in the figures. The plot on the left side of each GNR structure shows the average longitudinal e-flux, $j_x(E)$ at each dimer. The current profiles at the edges are similar for all the AGNRs. In the centre, three distinct patterns are obtained – (a) a clockwise ring, (b) an anti-clockwise ring, and (c) vanishing current. The schematic picture of these three patterns are illustrated in (e), (f), and (g) respectively. These three patterns repeat periodically with increasing $N_T$ with a period of $\Delta N_T = 3$. $r_c$ and W denote the cyclotron radius and the AGNR width, respectively. All computations were done at low energy such that there is effectively only one transport mode, i.e. only subband n=1 was included. (d) and (h) In the presence of a strong $B^\perp$-field, the spatial current profile is concentrated at the edges for all three types of AGNRs.

Firstly, we discuss the spatial distribution of electrons in an AGNR under an applied $B^\perp$-field. Figure 1 shows the spatial profile of the e-flux in AGNRs with different widths (total number of dimers, $N_T$=25,27, and 29) exposed to $B^\perp$-field under equilibrium condition. We set the intrinsic level, $E_i$ at the middle of the bandgap, to 0 for all the cases. The results were plotted at the Fermi level, $E_F - E_i$=0.3eV. The $E_F$ is located slightly above the conduction band ($E_C$) such that only one transport mode (n=1) is considered. Practically, $E_F$ can be adjusted by either doping the GNR [30] or applying back-gate voltage [18,20]. Figure 1a, 1b, and 1c show the spatial distribution of e-flux of AGNRs with $N_T$=25, 27, and 29, respectively. The e-flux between two atomic sites, $j_{mn}(E)$ are represented by the arrows in the figures. The plot on the left side of each GNR structure shows the average longitudinal e-flux, $j_x(E)$ for each dimer. For all the three types of AGNRs, the e-flux profiles at the edges are similar. At the top edge of the AGNRs, the average $j_x$ is negative while at the bottom edge the average $j_x$ is positive. In the bulk (central part of the AGNR), the average $j_x$ is zero. The variation of the average $j_x$ from negative to positive from top to bottom is caused by the external magnetic field, which redistributes the conduction electrons in the AGNRs to the edge, due to Hall effect. Under a finite $B^\perp$-field, the transverse location of the wavefunction is proportional to its velocity[17]. As the $B^\perp$-field increases, the forward ($X^+$) states, which carry the electrons in the forward direction shift to one side of the sample, while the backward ($X^-$) states, which carry the electrons in the backward direction shift to the other side of the sample [17]. Therefore, the $X^+$ and $X^-$ states are located at opposing edges of the structure and electrons are mainly transported only along the edges. For example in figure 1, the $X^+$ ($X^-$) states are located at the bottom edge (top edge).

Furthermore, our results show that the e-flux distribution profile, especially in the centre of the AGNR, is dependent on the width of the AGNR. There are three distinct e-flux profiles in the centre of the AGNR under $B^\perp$-field: 1) a clockwise ring (CR) pattern [figure 1(a)], 2) an anti-clockwise ring (aCR) pattern [figure 1(b)], and 3) vanishing current (VC) [figure 1(c)]. We find that these three profiles repeat periodically with $N_T$, with a periodicity of $\Delta N_T = 3$. i.e. when $N_T$=3p/3p+1/3p+2 [p is an integer number], the CR/ aCR/ VC pattern is obtained. This unique dependence of the spatial profile of e-flux on the AGNR width is attributed to the honeycomb lattice and the edge structure of the AGNR. Figure 1(e-g) illustrate how these three distinct e-flux patterns can arise at the centre of the honeycomb armchair structure with varying $N_T$. We label the AGNR as AGNR$^{3p}$, AGNR$^{3p+1}$, and AGNR$^{3p+2}$ corresponding to $N_T$=3p, 3p+1, and 3p+2, respectively. In all the AGNRs, the e-flux profiles at the edges are similar, i.e. negative (positive) $j_x$ at the top (bottom) edge. The profile of the top (bottom) edge is repeated along adjacent dimer rows, but with a decreasing intensity, as we proceed towards the bottom (top) of the AGNR. Thus at the upper half of the AGNR, the profile of the top edge dominates, while in the lower half of the AGNR the profile of the bottom edge dominates. At the center, the e-flux pattern is determined by the overlap of these two edge profiles. Depending on the width of the AGNR the overlapping pattern varies and thus the three unique profiles – CR, aCR, and VC – are obtained. Note that these three different AGNRs, i.e. AGNR$^{3p}$, AGNR$^{3p+1}$, and AGNR$^{3p+2}$, exactly correspond to the three different types of AGNRs classified based on the bandgap variations. [10]

Furthermore, due to its circulatory flow about each hexagon, the e-flux also oscillates between negative to positive values across the transverse width of the device, as shown by the $j_x$ plot. With increasing $B^\perp$-field, the amplitude of the $j_x$ oscillation increases (decreases) at the edge (bulk). When $\phi$ is high, such that the cyclotron radius of the electron, $r_c = S\sqrt{2Em}/(e\phi\phi_0) \propto \sqrt{E}/\phi$ is smaller than the AGNR width (i.e., $r_c < W$), the oscillation of $j_x$ in the bulk decreases to zero, and hence the difference between the three types of the AGNRs becomes insignificant. As shown in Figs. 1d and 1h, all three AGNRs will then have the same e-flux profile at high $B^\perp$-field, corresponding to $\phi$=0.05.

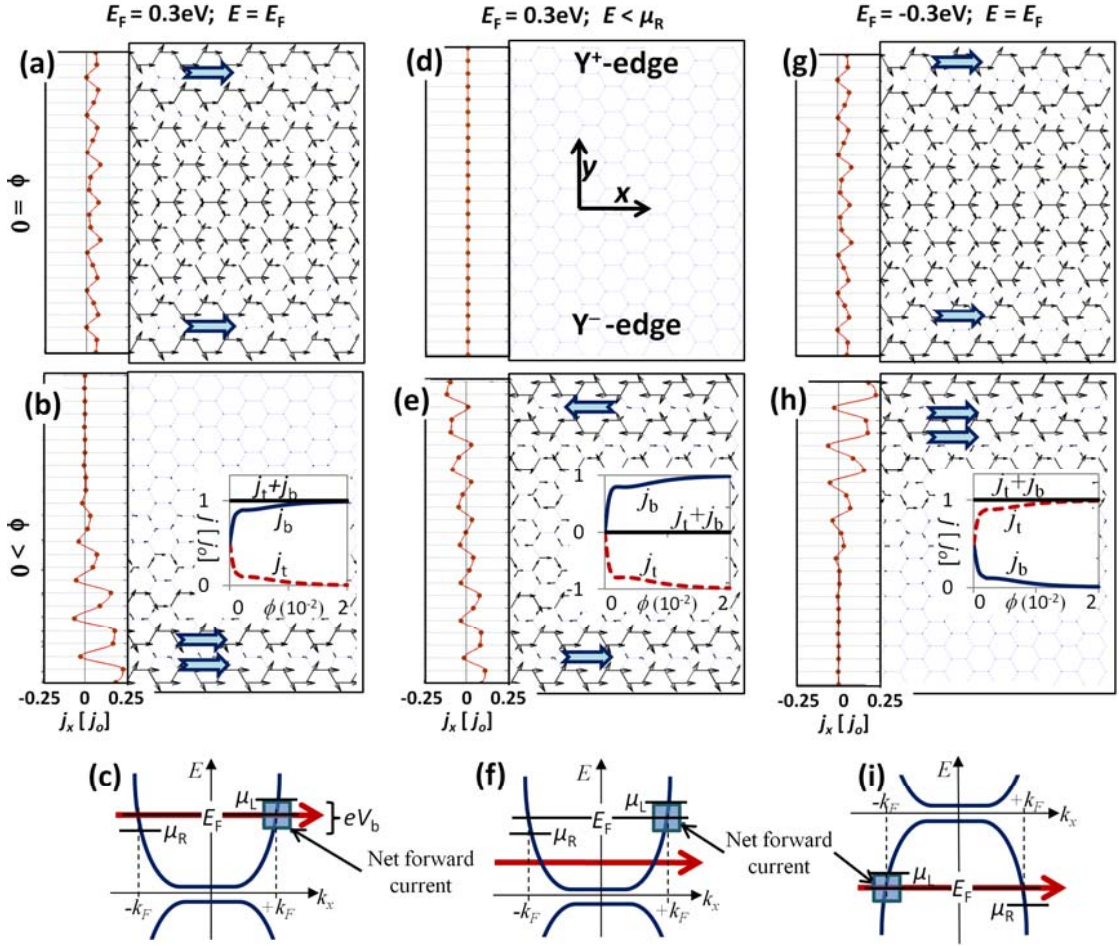

Figure 2 Spatial profile of j(E) under nonequilibrium condition (a) in the absence of $B^\perp$ field ($\phi$=0), and (b) in the presence of $B^\perp$ field corresponding to a flux of $\phi$=0.01. The inset in (b) plots the e-flux in the top (bottom)-half of the structure, $j_t$ ($j_b$), as a function of $\phi$. In (b) e-flux is deflected to the bottom of the AGNR due to magnetic field. The plot on the left of each structure shows the average longitudinal e-flux, $j_x$ across the AGNR. (c) E-k diagram in which the Fermi level is set in the conduction band of the AGNR, i.e. $E_F$=0.3eV. The arrow denotes the energy level at which the e-flux is computed in (b), i.e. $E = E_F$ = 0.3eV, while the shaded box denotes the states within the conduction window $\mu_L < E < \mu_R$, in which the forward e-flux is transported. (d)-(f) show similar plot as (a)-(c), but with the electron energy at which e-flux is computed being set below the conduction window, i.e., E=0.275eV < $\mu_R$. (g)-(i) shows similar plot as (a)-(d), but with the Fermi level $E_F$ in the valence band, i.e. $E_F$ = –0.3eV. All computations were done at low energy such that only one transport mode or sub-band, i.e. n=1, is included. We also assume a source-drain bias, $\mu_L$-$\mu_R$ = 0.01eV and temperature T=0 K.

Next, we analyze the spatial profile of the nonequilibrium e-flux in the AGNR in the presence of a $B^\perp$-field. Figure 2(a) shows the e-flux distribution under nonequilibrium condition, i.e. when a finite bias (0.01V) is applied across the structure. Under this condition, a net positive longitudinal e-flux is obtained across the AGNR [figure 2(b)]. When the Fermi level, $E_F$ is in the conduction band, the application of a $B^\perp$-field causes the nonequilibrium e-flux at $E=E_F$ to be deflected to the bottom of the structure as shown in figure 2(b). Referring to the inset of figure 2(b), at $E=E_F$ the e-flux across the top half ($Y^+$-edge) and bottom half ($Y^-$-edge) of the AGNR is given by $j_t$ and $j_b$, respectively. When $\phi=0$ the conductance at both edges are equal, i.e. $j_t = j_b = 0.5 j_o$ [figure 2(a)]. However, as $\phi$ increases, the conductance at one of the edges decreases to zero, i.e. $j_t \to 0$, whereas the conductance of the other edge increases, i.e. $j_b \to j_o$. However, the total e-flux remains constant, i.e. $j_b + j_t = j_o$, since only one transport mode (n=1) is considered in the computation, as shown in inset of figure 2(b).

The above may be explained by referring to the E-k diagram of figure 2(c), and noting that the velocity of the electron in state $k$ is proportional to the gradient of the E-k band, i.e $v_k \propto dE/dk$. As seen in the equilibrium case [figure 1(a)], the $B^\perp$-field tends to shift the $X^+$ ($X^-$) states with positive (negative) velocity to the $Y^-$-edge ($Y^+$-edge). In the E-k diagram of figure 2(c), we find that states with $k_x>0$ ($k_x<0$) has positive (negative) velocity, and will likewise be shifted to the $Y^-$-edge ($Y^+$-edge) under a $B^\perp$-field. Under nonequilibrium condition and zero temperature, the net current is transported within the conduction window, $\mu_L > E > \mu_R$, where $\mu_{L(R)}$ is the electrochemical potential of the left (right) contacts. Within this energy window, the Fermi distribution function on the left (right) contacts is one (zero), i.e. $f_L = 1$ ($f_R = 0$), so that the difference between the Fermi distribution functions, $\Delta f = f_L - f_R = +1$. As a result, net electron is transported only in the forward direction, i.e. $X^+$ electrons in the $Y^-$-edge [figure 2(b)].

For energies states below the conduction window, i.e. E< $\mu_R$, both $f_L$ and $f_R$ are 1, resulting in Δf=0. Therefore, these states are completely full and under zero $B^\perp$-field the net current is locally zero everywhere [figure 2(d)]. However, under finite $B^\perp$-field, both the $X^+$ and $X^-$ states are spatially separated at the edges of the structure, and thus the $X^+$ and $X^-$ e-fluxes do not cancel each other locally. As shown in figure 2(e) and its inset, both $X^+$ and $X^-$ e-fluxes with equal magnitude are obtained in the opposing edges, and as $\phi$ increases, the spatial separation of $X^+$ and $X^-$ states to the edges becomes more prominent. This increases the magnitude of the edge e-fluxes, resulting in circulating currents (depicted by arrows in figure 2(e)). Thus for E<$\mu_R$, unlike the situation under B=0, for finite $B^\perp$-field the local current along the edges is not zero. However, the overall e-flux integrated over the transverse direction remains zero due to Δf=0.

Next, we compare the electron motion under $B^\perp$-field when the Fermi level is located in the valence, as opposed to the conduction band. Interestingly, in the former ($E_F$ in the valence band), the net electron transport occurs across the $Y^+$-edge under finite $B^\perp$-field, whereas in the latter ($E_F$ in the conduction band), net electron transport occurs across the $Y^-$-edge, as shown in figure 2(h) and 2(b), respectively. Similarly, the e-fluxes in the $Y^+$-edge and $Y^-$-edge ($j_t$ and $j_b$) for the two cases show opposite dependence on $\phi$, as shown in the insets of figure 2(h) and Fig 2(b). This phenomenon may be explained based on the classical model by assigning a negative effective mass to electrons in the valence band, so that the Lorentz Force causes them to be deflected upwards, whilst electrons in the conduction band (which has a positive effective mass) are deflected downwards. However, the effective mass is not a parameter used in our simulations. Therefore, a proper explanation of this phenomenon has to be based directly on the E-k relation calculated by the tight binding model used in our simulation. Unlike the conduction band, the curvature of the valence band is negative and hence, the electrons with positive velocity in the valence band occupies the states with $k_x<0$. As a result, in the valence band, the $-k_F$ state which is dominated by $\mu_L$ is occupied, while the $+k_F$ state which is dominated by $\mu_R$ is empty. Thus, the forward electron transport occurs via the $-k_F$ state. [This is schematically depicted by the shaded box of figure 2(i)]. Furthermore, under $B^\perp$-field, the state with $-k_F$ is located at the $Y^+$-edge, thus resulting in the transport of Fermi electrons along the $Y^+$-edge.

In addition to the electron transport through a perfect AGNR structure, we also investigate the effect of backscattering by introducing atomic disorders in the structure. We consider atomic vacancies at three locations: (I) at the top-edge, (II) at the centre (bulk disorder), and (III) at the bottom-edge, of the AGNR, and these are represented by shaded rectangular boxes in figure 3. We find that without magnetic field [figure 3 (a)-(c)], i.e. $\phi=0$, the e-flux is higher across that disordered-edge for all the cases. This is because the disorders induce additional localized states [31-33] causing electron accumulation across the disordered edge. The conductance of the structure is also degraded, i.e. $g<g_o$, due to the scattering, i.e. $g=0.7g_o$ ($0.18g_o$) for edge (bulk) disorder. Note that all the simulations in figure 3 are done under single-mode (n=1) transport, and therefore, without any disorder the conductance $g=1g_o$.

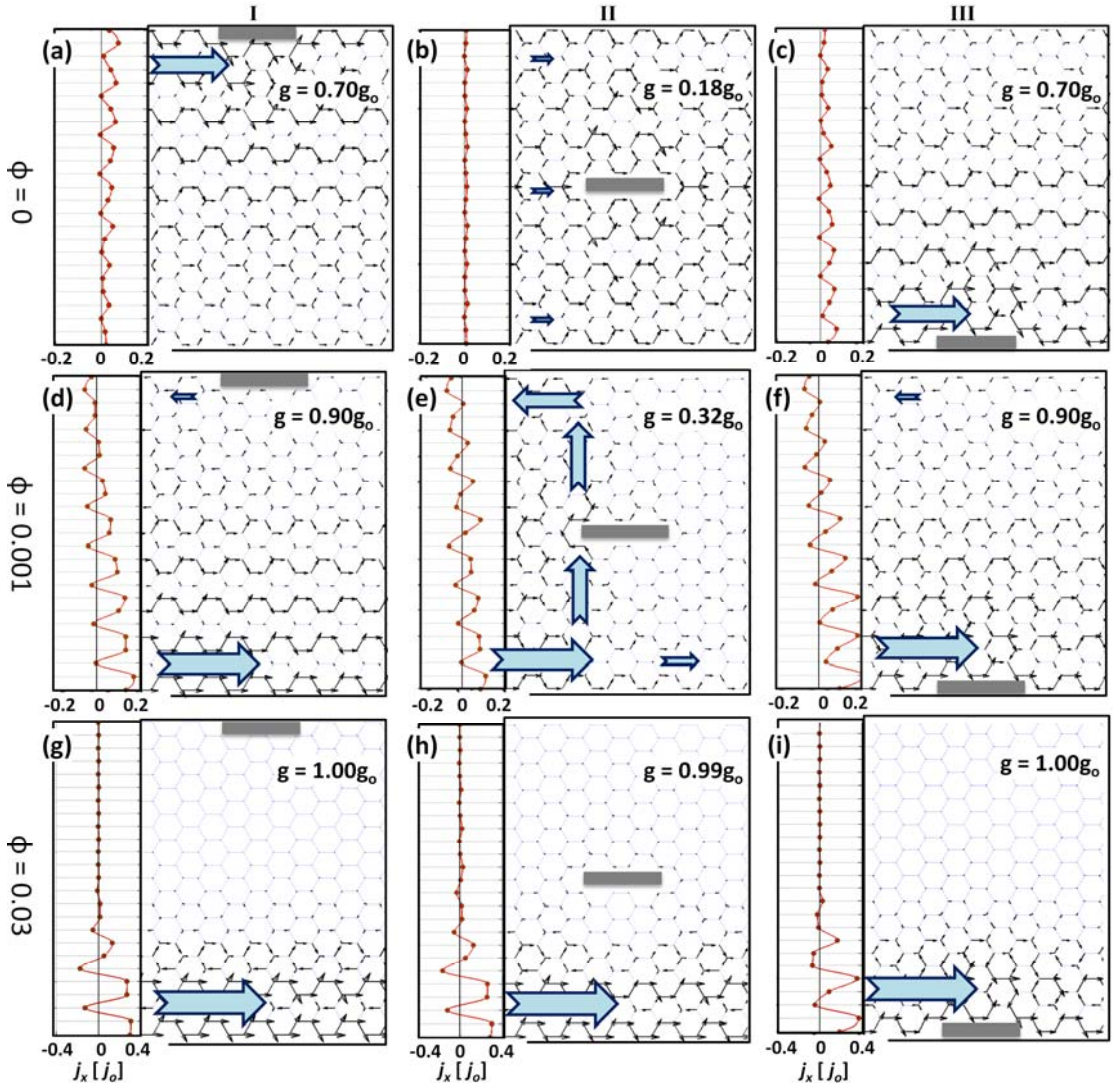

Figure 3 Spatial profile of the nonequilibrium e-flux, $j_{mn}(E)$ in AGNR structures with atomic disorders. Three types of disorders are shown: (I) vacancies at the top edge, (II) vacancies at the center, and (III) vacancies at the bottom edge. The vacancies are indicated by the shaded rectangular boxes. (a-c),(d-f), and (h-j) show the $j_{mn}(E)$ profile when $\phi=0$, 0.001, and 0.03. All computations were done at the Fermi level ($E_F$), and the $E_F$ was set to a low energy ($E_F=0.3eV$) such that only one transport mode, i.e. n=1 was included. The bias voltage is taken to be $V_b=0.01V$.

Figure 3d-f show the effect of moderate $B^\perp$-field ($\phi=0.001$) and disorders on spatial profile of e-flux. The field causes further confinement to the electrons, in addition to the confinement caused by the edges. Unlike Figure 3a, where the $X^+$ electrons mainly flow in the disordered $Y^+$-edge, Figure 3d shows that when the AGNR is exposed to a $B^\perp$-field, the $X^+$

electrons flow mainly in the Y$^-$-edge, even in the presence of disorders in the Y$^+$-edge. This is because, while the disorders induce localized states at the Y$^+$-edge, the B$^\perp$-field deflects the X$^+$ electrons to the Y$^-$-edge. When the disorders are located at the Y$^-$-edge [figure 3(f)], both the disorder induced states and the B$^\perp$-field deflect the electrons in the same direction, i.e. to the Y$^-$-edge.

The application of B$^\perp$-field enhances the conductance of the disordered structures. This is due to the confinement of the X$^+$ and X$^-$ states to the two edges, thus suppressing the backscattering effect. However, when the disorders are located at the center [figure3(e)], we find significant backscattering of electrons due to the bulk disorder induced intermediate states in the center of the AGNR. Under B$^\perp$-field, these intermediate states facilitate the backscattering of the X$^+$ electrons (at Y$^-$-edge) to X$^-$ states (at Y$^+$-edge). This results in a semicircular e-flux pattern to the left of the scattering center, as shown in figure 3(e). However, as B$^\perp$-field increases, the backscattering effects are suppressed since the electrons are more localized at edges, thus reducing the probability of being scattered by the disorder at the center. For example, when $\phi=0.03$ the cyclotron radius of electron is smaller than the AGNR width, leading to the quantum Hall Effect, where discrete Landau Levels are formed, and the transport states exist strictly only at the edges, resulting in $g\approx 1g_0$ for all disordered cases.

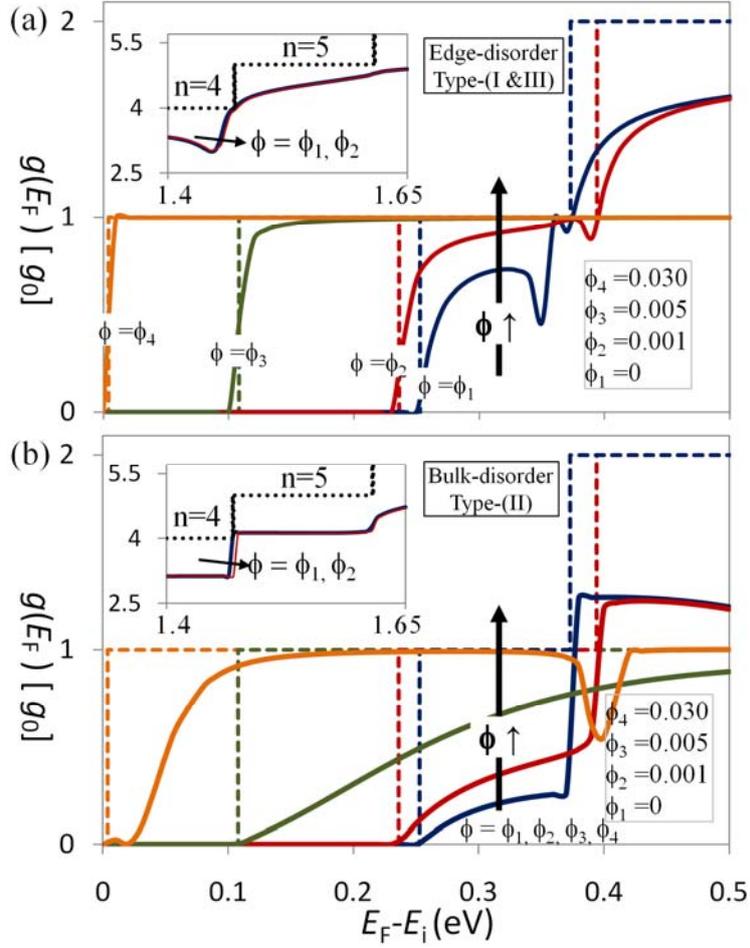

Figure 4 Conductance, $g(E_F)$ variation with fermi energy, $E_F$-$E_i$ ($E_i$=0) in AGNR with atomic disorders (a) in the edges (types I & III of figure 3), and (b) in the bulk (type II of figure 3). At low energy, when only one transport mode is included, i.e. n=1, the conductance is enhanced due to the application of perpendicular magnetic field ($B^\perp$-field). The inset shows the conductance g at higher energy when multiple transport modes are included, i.e. n=5. The effect of $B^\perp$-field is insignificant at higher energy when multiple transport modes are included, such that the conductance falls well below the ideal value of $g=ng_0$. As B-field increases, the transport gap decreases because the lowest sub-band $|n|=1$ shifts towards E=0 [12-16]. Furthermore, at certain energies, g(E) exhibits a non-monotonic behavior, i.e. suppression and oscillations [35,36], due to the vacancies in the GNR.

Figure 4(a) and figure 4(b) show the variation of conductance with varying $E_F$ for edge and bulk disordered AGNR, respectively. As discussed in figure 3, in general, 1) the conductance of the bulk disordered AGNR is much less than edge disordered AGNRs due to the backscattering effect, and 2) the conductance of a disordered AGNR is enhanced as $B^\perp$-field increases. However, when the Fermi level moves to the higher energy, more transport modes (n>1) can participate in the electron transport. For such multimode transport, the effect of $B^\perp$-field on the spatial distribution of electron and Hall Effect becomes less significant. This is because in multimode transport, where the E is high, the cyclotron radius of the electron $r_C \propto \sqrt{E}/\varphi$ is large. As a result, a higher ϕ is required to reduce the $r_C$ such that $r_C$<W, and shift the transport states to the edges. Therefore, the sensitivity of the electronic transport in AGNR to the $B^\perp$-field decreases. As shown in the inset of figure 4, the conductance at the higher energy (n=4 and n=5) degrades significantly in both the edge disordered and bulk disordered AGNRs even in the presence of a strong $B^\perp$-field.

Finally, we note that the vacancies may also affect the atomic bond structure of adjacent carbon atoms [34]. However in our tight-bonding model we do not consider the change in the bonding structure of the neighboring atoms. Although this may affect details of the transport properties, such as the magnitude of the conductance, the general performance of the structure would remain because the absence of back-scattering is a topological phenomenon, and is not dependent on the specific features of the defects.

## **Summary**

In summary, we have studied the electron transport in AGNR in equilibrium and nonequilibrium conditions under a finite $B^\perp$-field. In the equilibrium case, we identified three distinct spatial profiles of current corresponding to AGNR with $N_a$=3p/3p+1/3p+2, consistent with previous studies. These three different AGNRs, exactly correspond to the three different types of AGNRs classified based on the bandgap variations. The main part of our study focuses on the nonequilibrium current profile in AGNR. We consider transport at three different energy levels: 1) at Fermi level $E_F$ within the conduction band – where forward current is transported only along the bottom-edge, 2) within the conduction band but below the conduction window – where current is transported at both the edges in opposite directions with equal magnitude resulting in no net current, and 3) at $E_F$ within the valence band – where forward current is transported only along the top-edge. We also investigated the effect of atomic disorders and the induced backscattering on the current transport. We considered vacancy disorders in the bulk and at the edges of the AGNRs. Generally, the conductance of the AGNR is degraded due to the additional scattering caused by the disorders. However, we also found that the application of $B^\perp$-field could enhance the conductance in a disordered structure. This is due to the confinement of the forward and backward conduction states to the two edges, thus suppressing any backscattering effect. We further study this conductance enhancement by the $B^\perp$-field for multimode transport, which applies for high electron energies. Compared to the single mode transport, in multimode transport the conductance enhancement of the $B^\perp$ field is diminished, due to the larger $r_c$ at a given magnetic field, and hence weaker localization of the transport states along the edges. Finally, we would like to note all the above computations are performed on narrow AGNR structures with less than 30 dimers. This is to enable a clear graphical depiction of the electron flux profile. For wider AGNRs or even graphene sheets, similar effects should still be observed, and at lower (and more practical) $B^\perp$-field values.

**ACKNOWLEDGEMENT**

The computations were performed on the cluster of Computational Nanoelectronics and Nano-device Laboratory, National University of Singapore. This work was supported by Agency for Science, Technology and Research, Singapore (A*STAR) under grant number 082-101-0023.